\documentclass[english,pre,floats,superscriptaddress,usenames]{revtex4}
\usepackage[T1]{fontenc}
\usepackage[latin9]{inputenc}
\usepackage{amsmath}
\usepackage{amssymb}
\usepackage{graphicx}
\usepackage{esint}
\usepackage{babel}

\usepackage{amsmath}
\usepackage{epsfig}
\usepackage{graphicx}
\usepackage{float}

\usepackage[dvipsnames]{color}

\newcommand{\tr}[1]{{#1}}

\newcommand{\be}{\begin{equation}} 
\newcommand{\ee}{\end{equation}}
\newcommand{\bea}{\begin{eqnarray}}   
\newcommand{\eea}{\end{eqnarray}}

\newcommand{\ldeb}{{\lambda_{\mbox{\tiny D}}}}
\newcommand{\kdeb}{{k_{\mbox{\tiny D}}}}

\newcommand{\xx}{{\bf x}}
\newcommand{\rr}{{\bf r}}
\newcommand{\NN}{{\bf \nabla}}
\newcommand{\FF}{{\bf F}}

\newcommand{\vv}{{\bf v}}

\newcommand{\fa}{f^{\alpha}}
\newcommand{\na}{n^{\alpha}}

\newcommand{\cc}{{\bf c}}
\newcommand{\uu}{{\bf u}}
\newcommand{\uua}{{\bf u}^{\alpha}}

\newcommand{\uai}{u^{\alpha} _i}
\newcommand{\uaj}{u^{\alpha }_j}

\newcommand{\ma} {m}

\begin{document}

\title{Electro-osmotic flows under nanoconfinement: a self-consistent approach}

\author{Simone Melchionna}

\address{IPCF-CNR, Istituto Processi Chimico-Fisici, Consiglio Nazionale delle Ricerche, 
Universit\`a La Sapienza, P.le A. Moro 2, 00185 Rome, Italy}

\author{Umberto Marini Bettolo Marconi}

\address{Scuola di Scienze e Tecnologie, 
Universit\`a di Camerino, Via Madonna delle Carceri, 
62032 Camerino, INFN Perugia, Italy}

\selectlanguage{english}


\begin{abstract}
{
We introduce a theoretical and numerical method to investigate the properties of electro-osmotic flows under conditions of extreme
confinement. The present approach,  aiming to provide a simple modeling of electrolyte solutions described as
 ternary mixtures, which comprises two ionic species and a third uncharged component, is an extension of
our recent work on binary neutral mixtures. The approach, which  combines elements of kinetic theory, density functional theory
with Lattice-Boltzmann algorithms, is microscopic and  self-consistent and does not require the use of constitutive equations
to determine the fluxes. Numerical solutions are obtained by solving the resulting coupled equations for the one-particle
phase-space distributions of the species by means of a Lattice Boltzmann discretization procedure. Results are
given for the microscopic density and velocity profiles and for the volumetric and charge flow. 
}
\end{abstract}

\maketitle

 
 \section{Introduction}
The issue of transport of ionic solutions confined in nanometer scale channels
has been a concern of physicists, biologists and chemists for a long time due to their
importance in technological, biological and industrial applications.
With advances in nanofabrication techniques the study of fluid dynamics in atomically sized 
systems is now accessible and has become an active field of research \cite{kirby,bocquet}.

Electro-osmotic flow is an important phenomenon occurring in electrolytes when 
an electric field is applied along a channel and drags the electric double layer (EDL) 
that spontaneously forms next to the charged walls.
Due to the existence of Debye screening, the resulting velocity 
profile of the fluid strongly departs from the classical 
Poiseuille velocity distribution, being plug-like, with the consequence
of greatly reducing the viscous dissipation near
the channel walls. 
While the understanding of such phenomena in microchannels is almost complete,
in nanochannels the picture is less clear, because the density, charge and velocity profiles 
can be highly non-uniform at such reduced scales \cite{schoch}.

The basis to understand the transport of electrolytes  in narrow capillaries 
was laid down by Smoluchowski about a century ago \cite{bruus}.
This theory is a mesoscopic  treatment based on the combination of two continuous 
approaches: the formation of a non-uniform spatial charge near the charged boundaries, 
the EDL, is described via the Poisson-Boltzmann 
equation whereas the mass flow induced  by an applied electric field is determined
through the  Navier-Stokes equation.

In spite of their crude underlying assumptions, the Smoluchowski approach and its modern evolutions
capture most of the physics associated  with ion transport in microfluidic systems and represent a extremely useful 
workhorse in electrokinetic theory.
 On the scales characterizing the  microchannels  
\tr{($1 \mu$m to $100\mu$m)}, the equations of hydrodynamics together with the Poisson-Boltzmann 
equation describing the electric charge distribution provide a satisfactory picture 
of the interplay between diffusion of the various species, Coulomb interactions and 
mass transport. The coupled equations must be solved with the spatially nonuniform
boundary conditions needed to describe the presence of charges at the surface of the channel.  
Nevertheless, the validity of such an approach is questionable when the pore size becomes comparable
either with the size of the constituting molecules or with the typical length characterizing the
behavior of electrolytes, the Debye length $\ldeb\equiv1/\kdeb$.

A fully microscopic investigation of the system via atomistic simulation, 
such as Molecular Dynamics, is feasible, 
but while it provides an accurate description of the short-time, small-scale behavior,
due to the intrinsic noise present in simulations, it precludes to access large scales 
where macroscopic laws emerge and conditions of sub-millimolar concentrations
have to be considered \cite{Quiao}.

About forty years ago, Gross and Osterle (GO) formulated the very popular  space-charge model
where the electrostatic potential is determined by means of the Poisson-Boltzmann
equation, the diffusion of the charges by the Nernst-Planck equation and 
the velocity for the laminar flow by the Navier-Stokes equation \cite{Grossosterle}.
The approach proposed by Nilson  and Griffiths \cite{Nilson} partially goes beyond the ideal gas treatment of the ions of GO and 
introduces an explicit treatment of the solvent,
by employing  tools of  density functional theory to determine the  density of the three species, but
still considers the mean local velocity of the fluid at the macroscopic  Navier-Stokes level and the  transport coefficients
are not not determined self-consistently and microscopically,  but introduced as phenomenological parameters.  

In the present letter, we show that it is possible to determine within a unified framework
thermodynamic and structural properties and dissipative forces starting  from a
microscopic description of the interactions among the particles  in a simple 
model of ionic solution. 
For this purpose, a natural candidate to bridge the gap between macroscopic
and microscopic approaches which also treats on equal footing density and velocity inhomogeneities is
the modified Enskog-Boltzmann equation \cite{vanbeijeren}. There, the  complete information on the statistical description 
of a fluid is assumed to be encapsulated in the one-particle phase  space distribution function of each species present in the fluid.
From the knowledge of the distribution functions one computes the mean values of the 
physical observables of interest, such as charge, mass, velocity 
and temperature distributions.
Within this framework, hard-spheres represent a central reference to handle
memory-free collisional rules by means of a modified Enskog kernel \cite{umbmmelchionnalausanne}.

\section{Governing Equations}

Let us consider for the sake of simplicity a ternary mixture composed by hard spheres with a common diameter 
$\sigma$ and equal masses $m$. 
\tr{The model can be easily extended to include species of different diameters and masses.}
We associate positive and negative positive charges, $z_\pm$, to the species labelled $+$ and $-$, respectively, 
whereas the third component with label $0$ carries zero charge. 
The interactions between point charges and between these 
and the surface charges at the boundaries occur in a space characterized by a 
uniform dielectric constant $\epsilon$ \cite{oleskyhansencomment}. 

The electrostatic potential $\phi(\rr,t)$ acting on the charges at point $\rr$  is given 
by the contribution determined by the ionic charge distribution $\rho(\rr,t)$
via the Poisson equation:
\be
\nabla^2 \phi(\rr,t)= -\frac{\rho(\rr,t)}{\epsilon}
\ee
and to the "external" charges sitting on the electrodes. 
No retardation effects are included in our treatment.

To describe dynamically the system we associate to each species the one-particle 
distribution function $\fa(\rr,\vv,t) $ of each component $\alpha$, with
$\alpha=+, -, 0$,
whose evolution is governed by the following set of integro-differential equations:

\bea
&&
\frac{\partial}{\partial t}\fa(\rr,\vv,t) +\vv\cdot\NN \fa(\rr,\vv,t)
+\frac{\FF^{\alpha}(\rr)}{\ma}\cdot
\frac{\partial}{\partial \vv} \fa(\rr,\vv,t)
-\frac{e z_\alpha }{\ma}\NN \phi(\rr)\cdot
\frac{\partial}{\partial \vv} \fa(\rr,\vv,t)
=\nonumber\\
&&
-\omega[ \fa(\rr,\vv,t)- \psi^{\alpha}_{\perp}(\rr,\vv,t)]+\beta{\bf \Phi}^{\alpha}(\rr,t) 
\cdot(\vv-\uu(\rr,t)) \psi^{\alpha}(\rr,\vv,t) .
\nonumber\\
\label{breytwo}
\eea
The number density $\na(\rr,t)$ and the average velocity $\uu^\alpha$ of species $\alpha$ are obtained
 by multiplying by $1$ and $\vv$, respectively, the distribution  $\fa$ and integrating w.r.t. $\vv$. 
The average packing fraction is $\xi=\pi/6\sigma^3 \sum_\alpha \na$ and the local barycentric velocity 
of the composite fluid is:
\be
\uu(\rr,t)=\frac{\sum_{\alpha}\na(\rr,t)\uua(\rr,t)}
{\sum_{\alpha}\na(\rr,t)}.
\ee

The temperature $k_BT\equiv1/\beta$,
with $k_B$ being the Boltzmann constant, is assumed to be uniform throughout the system and
each species has thermal velocity $v_T=\sqrt{k_BT/m}$.
In addition, $\FF^\alpha$  and ${\bf \Phi}^{\alpha}$ represent the sum of external and internal 
forces respectively of non-electrostatic nature acting on species $\alpha$. 
It should be noted  that electrostatic interactions are introduced at 
mean-field Vlasov level and due to the nature of the approximation involved do not contribute to the dissipative
character of the fluid, while they contribute to the chemical potential \cite{JCP2011}.
\tr{As discussed in ref. \cite{oleskyhansen}, adding a Coulomb correlation to the local
chemical potential does not significantly alter the
ionic and solvent densities as compared to the mean field approximation for electrostatics.}

The l.h.s. of eq. (\ref{breytwo}) encodes the effect of streaming on species $\alpha$ while
the r.h.s. contains the sum of two terms, a first one being a modified Bhatnagar-Gross-Krook relaxation 
kernel towards a Maxwellian \cite{BGK,JCP2011}, with
\be
\psi^{\alpha}(\rr,\vv,t)=\na(\rr,t)\frac{1}{(2\pi v^2_T)^{3/2}}\exp
\Bigl(-\frac{(\vv-\uu(\rr,t))^2}{2 v^2_T} \Bigl)
\ee
and
\bea
&&
\psi^{\alpha}_{\perp}(\rr,\vv,t)=\psi^{\alpha}(\rr,\vv,t) \Bigl\{1+
\frac{(\uua(\rr,t)-\uu(\rr,t))\cdot(\vv-\uu(\rr,t))}{v^2_T}\Bigl\} 
\nonumber\\
\label{prefactor}
\eea
where the collision frequency $\nu$ relates to the ideal contribution to viscosity.
\tr{To summarize the meaning of eqs. (\ref{breytwo}-\ref{prefactor}), the viscosity of the system must 
contain two contributions: the ideal viscosity, that we represent by the BGK term, 
and the hard-sphere viscosity, that we represent by the last
term of eq. (\ref{breytwo}). This second contribution is strongly modulated by the fluctuations 
in density and velocity, whereas the BGK term is not.
As far as the meaning of the baricentric velocity at atomic scale is concerned, one 
can consult the work of Travis and Gubbins \cite{travisgubbins}.}

The second term in the r.h.s. of eq. (\ref{breytwo})  is a collisional contribution. Since all interactions, but the 
hard-core repulsion are treated within the mean-field approximation, the form of ${\bf \Phi}^{\alpha}$ solely depends
on the properties of the underlying hard-sphere mixture. Therefore, we can use the results
already established for neutral systems and derived from a simplified treatment
of the Revised Enskog Theory (RET) \cite{vanbeijeren} due to Santos et al.  \cite{Brey}.



Deriving the mesoscopic equations for the model above is a relatively simple task
and can be achieved by using the same methods presented in refs. \cite{Melchionna2008-2009,JCP2011}
One obtains an equation for the charge distribution, an equation for the velocity distribution, 
assumed to be the same for all components.
The continuity equation reads
\be
\frac{\partial}{\partial t}\na(\rr,t) +\nabla\cdot \Bigl(\na(\rr,t) \uu(\rr,t)\Bigl)+
\nabla\cdot \Bigl(\na(\rr,t)(\uua(\rr,t)- \uu(\rr,t)\Bigl)=0 ,
\label{continuity}
\ee

The propagation of momentum of the single species obeys the following equation
\tr{
\bea
&&
\frac{\partial}{\partial t}[\na(\rr,t)\uaj(\rr,t)]+
\frac{\partial}{\partial x_i} \Bigl(\na(\rr,t) \uai(\rr,t) \uaj(\rr,t)
-  \na(\rr,t)(u^{\alpha}_i(\rr,t)-u_i(\rr,t))( u^{\alpha}_j(\rr,t)-u_j(\rr,t))\Bigl)=
\nonumber\\
&&
- \frac{\partial}{\partial x_i} \frac{ \pi_{ij}^{\alpha}(\rr,t)}{\ma}+ \frac{F^{\alpha}_j(\rr)}
{\ma}\na(\rr,t)+
 \frac{ \Phi^{\alpha}_{j}(\rr,t)}{\ma}\na(\rr,t) -\frac{e z_\alpha}{\ma}\na(\rr,t) \frac{\partial}{\partial x_j}\phi(\rr,t)
\label{momentcomponent}
\eea
}
where
\be
\pi_{ij}^{\alpha}(\rr,t)=\ma\int d\vv (v_i-u_i)(v_j-u_j)\fa(\rr,\vv,t)
\label{pressurekin}
\ee
is the kinetic contribution of component $\alpha$ to the pressure tensor.
As a result of a detailed microscopic calculation one can show that the  
average local force acting on a particle of  species $\alpha$ due to the influence of the remaining
particles can be subdivided into  three different contributions of non electrostatic nature,
associated with the excess chemical potential, the drag and viscous forces:
 \be
{\bf \Phi}^{\alpha}(\rr,t)= \FF^{\alpha,mf}(\rr,t)+\FF^{\alpha,drag}(\rr,t)+\FF^{\alpha,visc}(\rr,t) .
\label{splitforce}
\ee
The explicit form for these quantities depends on the microscopic details and for the charged hard-sphere
model employed expressions can be found in ref. \cite{JCP2011}.


\section{Results}

The numerical solution of eq. (\ref{breytwo}) is obtained by a generalization of the method 
presented in ref. \cite{Melchionna2008-2009} to the case of a ternary mixture, 
\tr{following the ideas outlined in ref. \cite{Moroni}}.
Similarly to the Lattice Boltzmann method of solving the BGK dynamics \cite{LBgeneral,Melchionnasucci}, 
our numerical approach spatially discretizes the distribution function over a cartesian mesh. 
The velocity dependence of the distribution is represented according to a truncated Hermite 
expansion supplemented by a Gauss-Hermite quadrature to evaluate the kinetic moments
\cite{shanyuanchen}.
The end result is a replacement of the distribution function with a discretized version,
$\fa(\rr,\vv,t) \rightarrow \fa_p (\rr,t)$, where the index $p$ labels a set of discretized 
velocities $\{\cc_p\}$.
In the following, we use the so-called D3Q19 mesh, including $19$ discrete speeds related 
to exchanges of particles between zeroth, first and second mesh neighbors. 
The Hermite expansion introduces a characteristic speed that we identify with the 
thermal speed $v_T$, being the same for each fluid component.

The numerical accuracy of the simulations is controlled by the truncation 
level of the Hermite expansion and the associated Gauss-Hermite quadratures in velocity space.
The collisional contribution in eq. (\ref{breytwo}) 
and the Maxwellian in the BGK-like term are expressed up to second order Hermite components.  
The number density and fluid velocity are obtained as 
$n^\alpha=\sum_p \fa_p$ and $n^\alpha \uu^\alpha=\sum_p \cc_p \fa_p$, respectively.
Finally, external forces are implemented according to the method suggested in ref. \cite{Guo}
in order to achieve second-order accuracy in time.

The method employed to solve the ternary system is further complemented with the solution
of an auxiliary distribution to minimize the effect of parassitic currents in the system.
In fact, given the strong density gradients in the system, the LBM method has the attitude to 
develop spurious currents in proximity of confining walls, where
the density of species has the strongest variations. \tr{This limitation is
intrinsic to the spatial discretization of the Lattice Boltzmann method. 
To solve this issue we have developed
an {\em ad-hoc} strategy to circumvent such problem, that will be presented 
elsewhere \cite{JCPpreparation}. The main idea is to use an implicit trapezoidal rule
for the time integration together with a version of the two-distribution method
of He {\em et al.} \cite{hechenzhang}.}

The systems are simulated in three-dimensions in slit geometries of variable width $W$ along the $z$
direction, with $-W_{eff}/2<z<+W_{eff}/2$ where $W_{eff}=W-\sigma$ is the available width,
and being periodic in the $x$ and $y$ directions.
\tr{To resolve accurately the density profiles, we have chosen a hard sphere diameter to
be equal to $8$ in lattice spacing units.}
At positions $z=\pm W_{eff}$, no-slip boundary conditions are imposed for the fluid velocities
via the bounce-back method applied on the populations of each species \cite{LBgeneral} and
at this level, we disregard slippage effects together with the distinction between the Stern layer 
and the shear layer. We consider a 1:1 symmetric electrolyte by enforcing electroneutrality
on the composite saline solution and charged surface system. The Poisson equation is solved numerically at
every  iteration of the LB algorithm by a successive over-relaxation (SOR) method. Second-order accurate, 
von Neumann boundary conditions on the electric field are imposed, that is,
$\hat n\cdot \partial \phi|_{z=\pm W} = -\Sigma/\epsilon$, where $\Sigma$ is the superficial charge density
on the plates. The Bjerrum length, $l_B=e^2/(4\pi\epsilon k_B T)$, is taken equal to
$l_B = 1.5 \sigma$, a value compatible with bulk water conditions at ambient temperature,
and data are collected for a uniform electric field $n \sigma^3 eE/m\eta_{id}^2 = 0.005$ 
acting in the direction parallel to the walls.

\tr{
We remark that the mean free path of a hard sphere system in bulk is
$\lambda_{mf}=\frac{1}{\sqrt 2 \pi g(\sigma)  \sigma^2 n}$.
By confining the system in a slit of width $W$,
the Knudsen number expresses  the ratio of the two
characteristic lengths of the problem, that is, 
$Kn =\frac{  \lambda_{mf} }{W} \simeq 0.1$ for the systems that we consider in this letter.
In other words, collisions are frequent enough to restore local equilibrium.}

\begin{figure}[h]
\begin{centering}
\includegraphics[width=7.5cm,clip=true,angle=0]{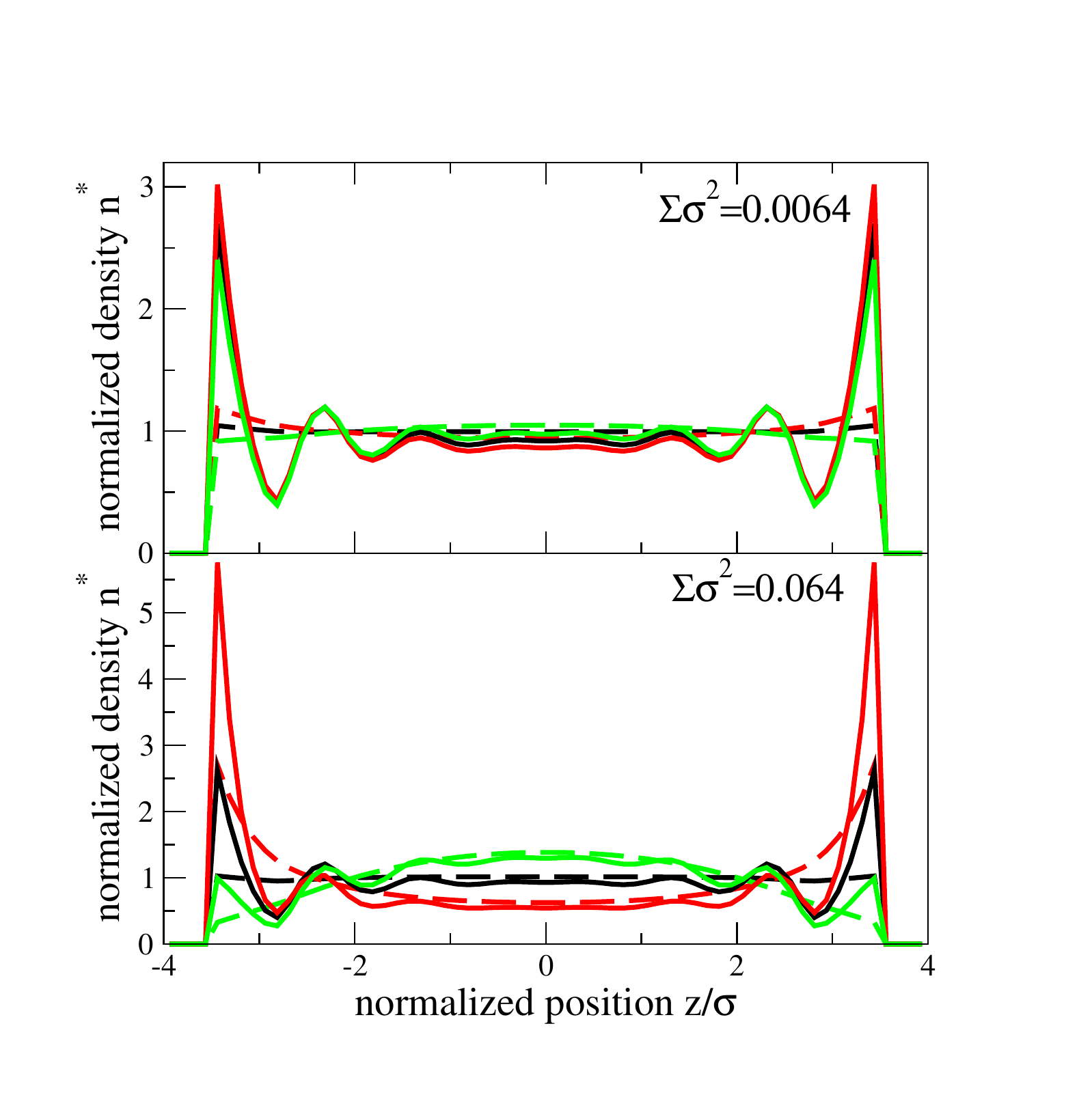}
\includegraphics[width=7.5cm,clip=true,angle=0]{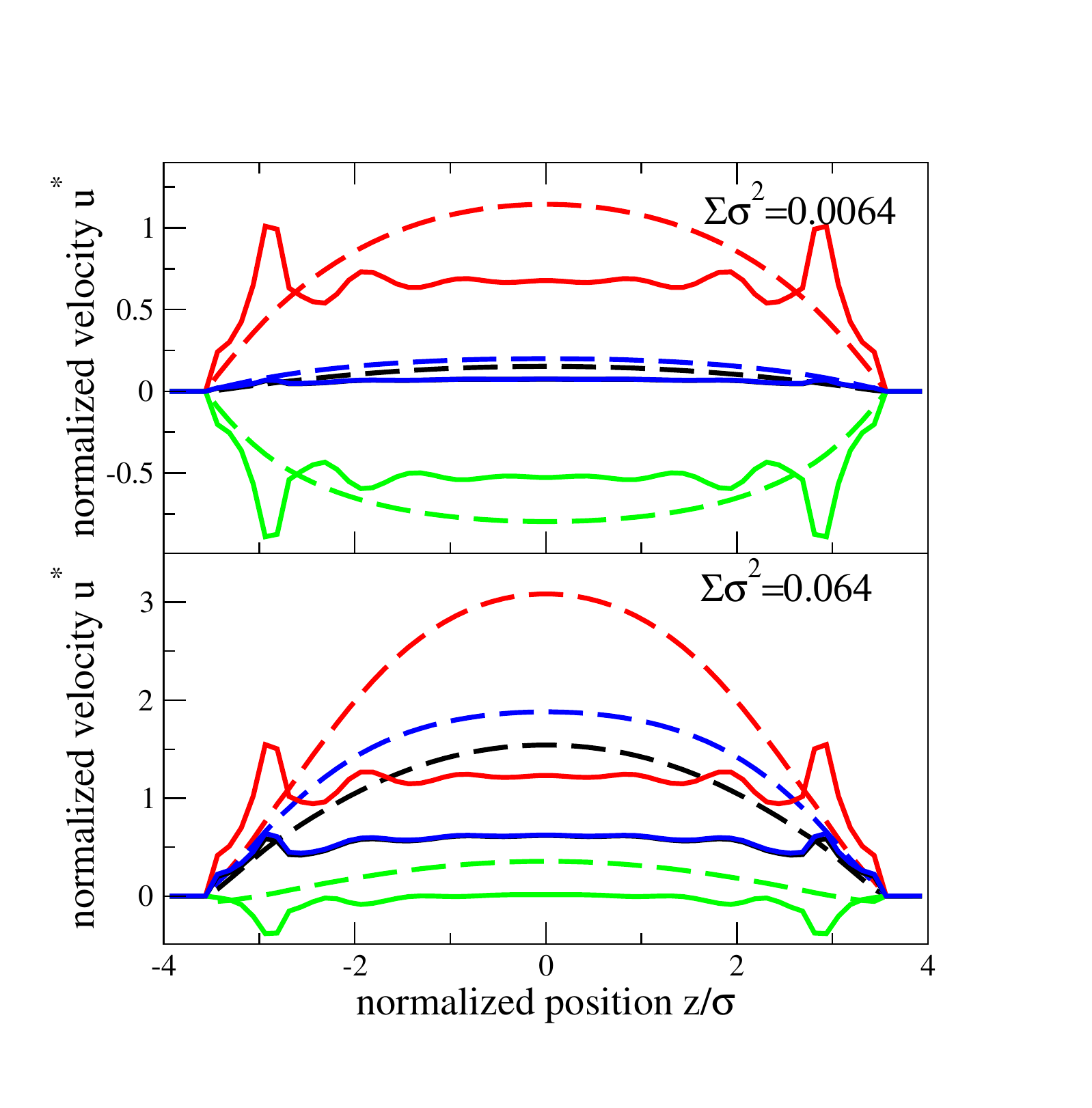}
\end{centering}
\caption{Profiles of solvent (black), counterions (red) and coions (green) for $0.28$ (solid lines) and 
virtually zero (dashed lines) packing fractions.
The left and right panels correspond to normalized density profiles 
and normalized velocity profiles, respectively. 
The upper and lower panels correspond to surface charge $\Sigma\sigma^2/e=0.0064$
and $\Sigma\sigma^2/e=0.064$, respectively. The data are for a slit with $W=64$ and $\kdeb W=5.33$}
\label{fig:profiles}
\end{figure}

The density and velocity profiles for two values of the surface charge are shown in 
Fig. \ref{fig:profiles}.
The profiles are obtained for packing fraction $\xi=0.28$  and are compared with the case of 
negligible packing fraction, whose evolution corresponds to electrokinetic motion in the absence 
of steric forces.
The reported normalized densities are $n^{\alpha,*} = n^\alpha / n_0^\alpha $ where $n_0^\alpha$ is 
the corresponding bulk value, whereas the normalized velocities are 
$u^{\alpha,*}=u^\alpha/u_0$, with $u_0=eE/(\eta_{id} 4\pi l_B)$ and
$\eta_{id}$ being the ideal contribution to the dynamic viscosity of the mixture.

The data in Fig. \ref{fig:profiles} demonstrate the strong oscillations in the density in 
proximity of the walls and the large differences with the case of the ideal electrolyte, 
with the characteristic cusp of the hard-sphere fluid at contact with the wall. 
The oscillations are further modulated by the development of the Debye layer, whose strength
increases with the surface charge. Analogously, the velocity profiles develop non-trivial
shapes that are far from quasi-parabolic, as compared to the ideal solution.
It is worth noticing both the peaks in the velocities of the charged species 
close to the wall, and the plug-like shape in the inner part of the slit. 
The inhomogeneities in the velocity profiles denote major departures from the
ideal solution case with a reduction of the velocities in the bulk region and
an overshooting near the walls. As the figure reveals, as a general trend
the barycentric velocities have a profile that stays closer to that of the neutral species than 
in the ideal case due to a drag effect.

\begin{figure}[h]
\begin{centering}
\includegraphics[width=7.5cm,clip=true,angle=0]{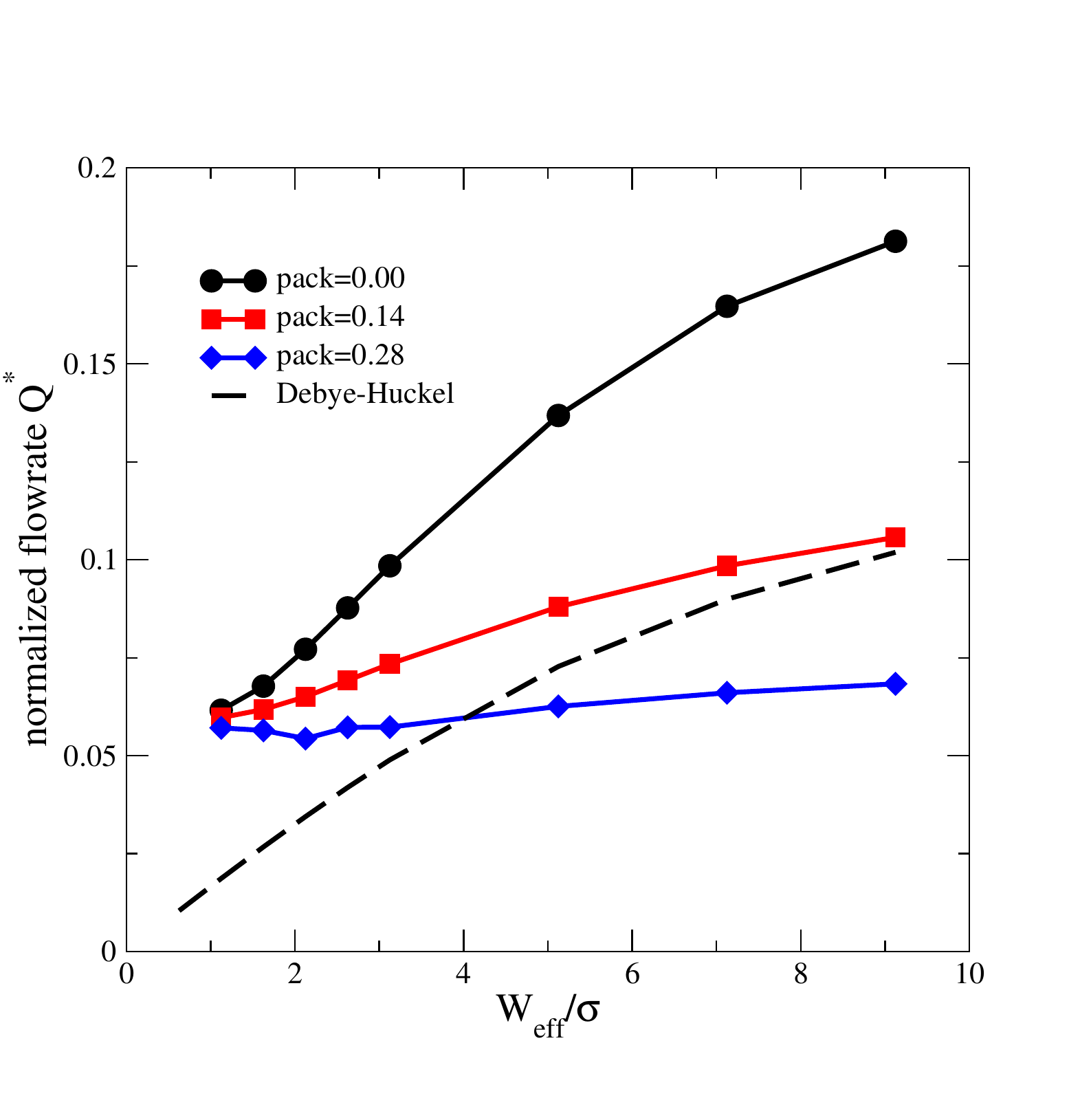}
\includegraphics[width=7.5cm,clip=true,angle=0]{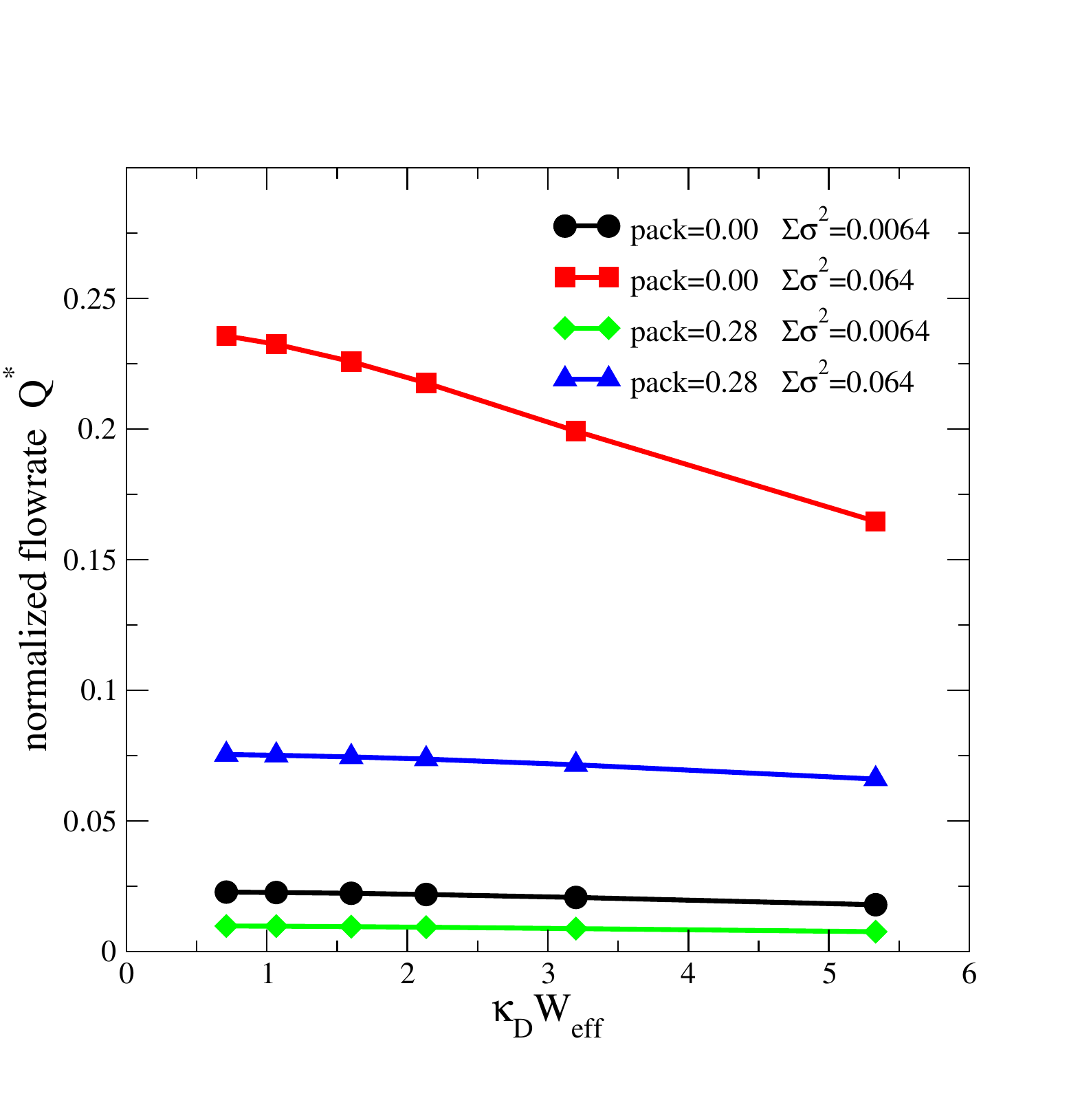}
\end{centering}
\caption{Normalized flowrates obtained for varying slit width $W_{eff}$ at fixed $\kdeb =1/1.5 \sigma$ (left panel) and 
varying $\kdeb $ at fixed $W_{eff}/\sigma=8$ (right panel). 
Data in the left panel are collected for different packing fractions ($\sim0.0$ for circle, 
$0.14$ for square and $0.28$ for diamond symbols). 
The Debye-H\"uckel prediction in a equivalent channel width $W_{eff}$ is reported as a dashed line.
Data in the right panel are collected for $\Sigma \sigma^2/e=0.0064$ and for 
different packing fractions ($\sim0.0$ for circle, and $0.28$ for diamond symbols) and 
for $\Sigma \sigma^2/e=0.064$ and for different packing fractions 
($\sim0.0$ for square, and $0.28$ for triangle symbols).
}
\label{fig:flowrates}
\end{figure}

The competition of the three species arising from drag and viscous forces produces the net 
flowrates shown in Fig. \ref{fig:flowrates}. Data are reported for different pore widths 
and for different values of the Debye length, ranging from thin layer ($\ldeb \ll W_{eff}$) 
to overlapping Debye layer conditions ($\ldeb > W_{eff}$). The normalized flowrates, 
computed as $Q^* \equiv \frac{1}{W_{eff}} \int dz u^*_z(\xx)$, are compared with 
the prediction arising from the Debye-H\"uckel (DH) level of theory, being
\be
Q^*_{DH}=\frac{\Sigma e}{k_BT\epsilon\kdeb} 
\left[\frac{1}{\tanh(\kdeb W_{eff}/2)}
 - \frac{1}{\kdeb W_{eff}/2}\right]
\label{DHflowrate}
\ee
The data show large departures from the DH prediction, with a strong suppression 
of flowrates at increasing slit width due to steric effects.
\tr{The results of Fig. \ref{fig:flowrates} demonstrate the non-trivial effects
of nanometric crowding, where the flowrates decrease as the packing fraction increases
with a strong dependence on the degree of confinement.}

\begin{figure}[h]
\begin{centering}
\includegraphics[width=12cm,clip=true,angle=0]{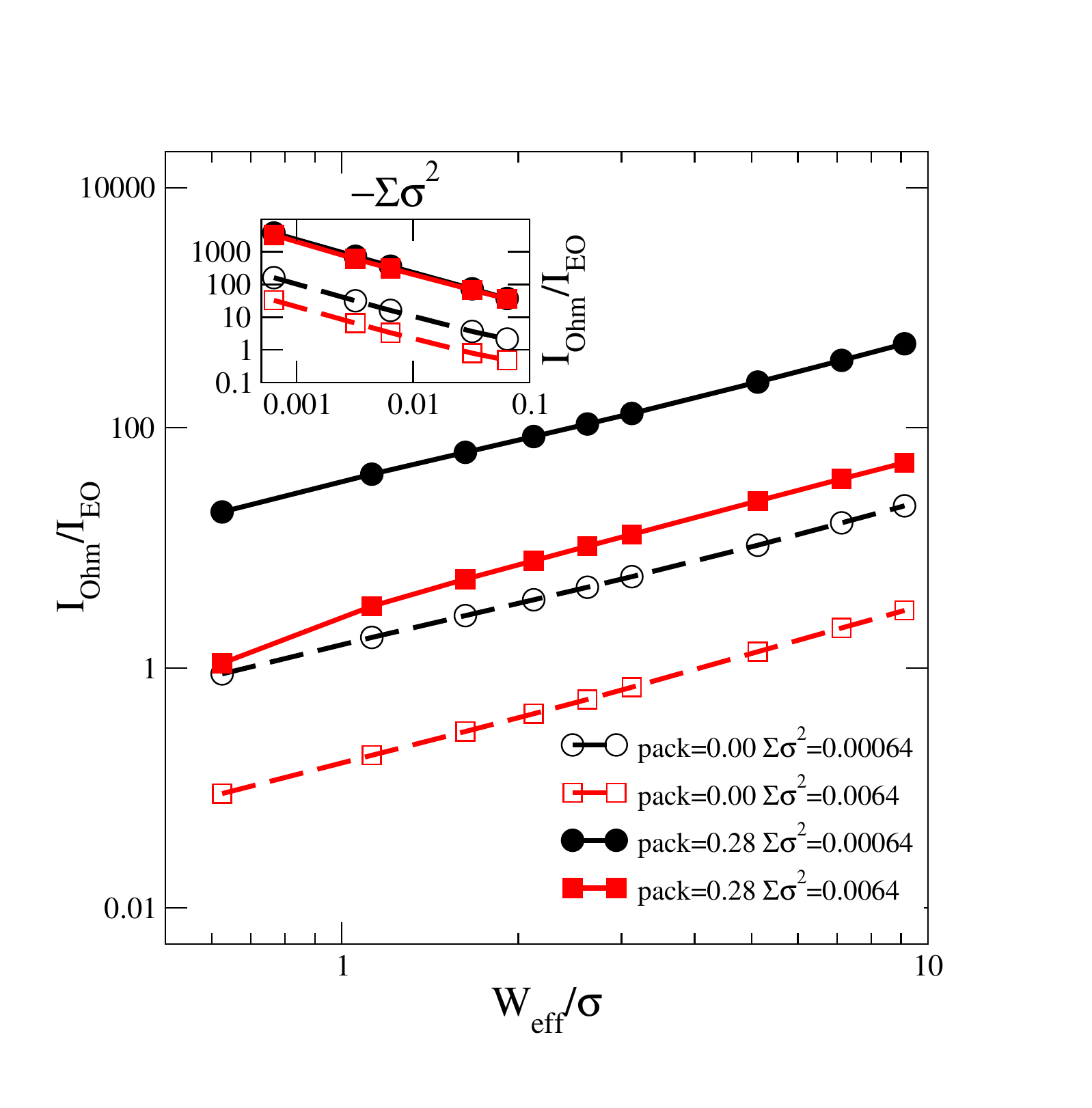}
\end{centering}
\caption{Log-log plot of the ratio of electrophoretic vs electroosmotic components to the ionic current 
as a function of the slit width.  Data are obtained for 
packing fraction $\sim 0.0$ (open symbols) and  $0.28$ (filled symbols) and for surface charge
$\Sigma \sigma^2/e=0.0064$ (circles) and $0.00064$ (squares). Inset: the same ratio reported as a function of
the surface charge and for $W/\sigma=8$.}
\label{fig:currentratio}
\end{figure}

In nanofluidic applications, it is important to evaluate the role of steric interactions
in evaluating the balance of bulk dominated versus surface dominated charge current.
We then compute the electro-osmotic charge current,
\be
I_{EO}=\int^{W_{eff}}_{- W_{eff}} (n^+(\xx) - n^-(\xx)) u_z(\xx)
\ee
which is of convective origin, and compare it to the conductive component, $I_{Ohm} = I - I_{EO}$, 
where the total current is measured as
\be
I=\int^{W_{eff}}_{- W_{eff}}  dz (n^+(\xx)u_z^+(\xx) - n^-(\xx)u_z^-(\xx))
\ee
Fig. \ref{fig:currentratio} reports the ratio of currents and shows a large dominance of the Ohmic 
component stemming from steric interactions for all values of the slit geometry and for two values
of the surface charge considered, as compared to the virtualy zero packing fraction case. These
results provide important indications on the effect of ionic crowding in nanospaces, as
much as providing a guide to experiments and theoretical investigations in nanofluidic set-ups. 


\section{Conclusions}

In summary, the results presented in this letter provide a working framework for
applying theoretical analysis and simulation to nanofluids. 
The method delivers unique informations relative to the transport of ionic solutions 
confined at molecular scale. 
The theory presented in this letter is self-consistent in such a way as to include
thermodynamic, drag and viscous collisional forces in a unified way and avoiding the
use of fudge parameters. The simulation is stable and robust, as it delivers
the noise-free distribution of the hydrodynamic moments for a wide range of parameters.
Work is in progress to analyze systematically the several consequences
of collisional forces in electrokinetics.



\begin{thebibliography}{99}
\bibitem{oleskyhansencomment}
\tr{In principle, to describe the polarizability of the medium
one could consider the zero component of the mixture as carrying a permanent dipole,
see e.g. \cite{oleskyhansen}.}
\end{thebibliography}
\end{document}